# An Artificial Intelligence (AI) workflow for catalyst design and optimization

Nung Siong Lai[1], Yi Shen Tew[1], Xialin Zhong[1], Jun Yin[2], Jiali Li[2], Binhang Yan[1], Xiaonan Wang[1]*

[1] Department of Chemical Engineering, Tsinghua University, Beijing 100084, China

[2] Department of Chemical and Biomolecular Engineering, National University of Singapore, 117576, Singapore

*Email: wangxiaonan@tsinghua.edu.cn   Tel: +86 10 62784572*

## Abstract

In the pursuit of novel catalyst development to address pressing environmental concerns and energy demand, conventional design and optimization methods often fall short due to the complexity and vastness of the catalyst parameter space. The advent of Machine Learning (ML) has ushered in a new era in the field of catalyst optimization, offering potential solutions to the shortcomings of traditional techniques. However, existing methods fail to effectively harness the wealth of information contained within the burgeoning body of scientific literature on catalyst synthesis. To address this gap, this study proposes an innovative Artificial Intelligence (AI) workflow that integrates Large Language Models (LLMs), Bayesian optimization, and an active learning loop to expedite and enhance catalyst optimization. Our methodology combines advanced language understanding with robust optimization strategies, effectively translating knowledge extracted from diverse literature into actionable parameters for practical experimentation and optimization. In this article, we demonstrate the application of this AI workflow in the optimization of catalyst synthesis for ammonia production. The results underscore the workflow's ability to streamline the catalyst development process, offering a swift, resource-efficient, and high-precision alternative to conventional methods.

## 1. Introduction

The development of novel catalysts to address increasing energy demand and consumption has become an urgent task in the realm of renewable energy [1]. The impetus for this development is illustrated by the growth in the global catalyst market, which was valued at USD 29.7 billion in 2022 and is projected to grow at a compound annual growth rate (CAGR) of 4.6% from 2023 to 2030. This surge is driven not only by escalating demands from applications in process optimization, yield improvement, and energy saving but also by a heightened awareness and concern for environmental issues, particularly the increase in carbon dioxide emissions. The proposition of carbon neutrality has illustrated the transformation path for the chemical industry over the coming decades, and set forth new challenges regarding the use of renewable energy and the catalytic conversion of carbon dioxide into high-value chemical products [2, 3]. Undeniably, the development of novel catalysts is crucial in addressing our energy needs and environmental concerns, yet it presents an arduous task, owing to the multifaceted nature of the problem at hand [4].

The path to new catalyst development is beset with three salient challenges [5]. First, the traditional catalyst development process, especially in the context of solid catalysts, requires extensive preliminary efforts. The sequence of catalyst synthesis, activity testing, characterization, and industrial scale-up form a complex and time-consuming process [6, 7]. Identifying optimal synthesis methods and process parameters requires extensive experimental data. The second challenge arises from the limitations imposed by time and material resources [7]. These constraints restrict chemists to exploring only the tip of the iceberg of the vast, high-dimensional chemical parameter space, thereby leaving potentially superior catalysts undiscovered [8]. The complexity of navigating this parameter space is underscored in the multi-step catalyst development process, where the crucial interrelation between catalytic activity and process variables is often neglected [9, 10]. Lastly, the advancement of parallel screening and high-throughput experimental technologies, while promising, presents its challenges. These methods, though more efficient than traditional trial-and-error techniques, demand greater resources in terms of cost and data volume [11], while given the lack of extensive data on new catalysts in lab conditions, the importance of small-scale data optimization is further highlighted [12]. The rapid synthesis of catalysts under various conditions, subsequent performance-based screening and optimization, and catalyst characterization all contribute to these escalating demands.

Several optimization strategies are conventionally employed to identify the optimal set of condition parameters, thereby enhancing the performance of the catalyst. The 'One Factor At a Time' (OFAT) method is frequently employed as an alternative technique for chemical process optimization and comprehension [13]. The OFAT approach often misinterprets chemical processes due to its inefficiency, inaccuracy, and neglect of synergistic effects and nonlinear responses among experimental variables [14]. Design of Experiments (DoE), a robust and extensively utilized optimization technique especially within the pharmaceutical and fine chemical sectors, serves to address these shortcomings [15]. DoE comprises a suite of statistical methodologies that endeavor to construct a mathematical model capable of describing the output of a chemical reaction (e.g., reaction yield, purity, etc.) based on the experimental inputs (e.g., temperature, reaction time) associated with the reaction. While these conventional optimization methods and their advancements have undeniably made significant contributions to the field, certain gaps persist that limit their full potential in optimizing catalyst synthesis. The predominant reliance on the empirical knowledge and intuition of seasoned chemists, while invaluable, is not systematically scalable and transferable. Techniques like OFAT and DoE, though statistically rigorous, are often unable to keep pace with the sheer complexity and vastness of the catalyst parameter space, leaving much of it unexplored and underutilized. These methods can struggle to account for the nuanced interdependencies among experimental variables and the nonlinear nature of chemical reactions [16, 17].

With the advent of machine learning (ML), the field of catalyst optimization is entering a new era. ML techniques are gaining traction due to their capability to model complex, nonlinear systems and find correlations in vast datasets that might otherwise be overlooked [18, 19]. Active learning based on Bayesian optimization (BO) [20], a type of ML, presents a promising solution to the shortcomings of conventional optimization techniques. This strategy is based on constructing a probabilistic model for the objective function and then using it to select the most promising parameters to evaluate in the actual objective function, based on a balance between exploration and exploitation [21]. This allows for more efficient sampling of the parameter space compared to OFAT and DoE, which can accelerate the optimization process and potentially discover better solutions. In a recent study, Rossmeisl et al. demonstrated a novel application of Bayesian optimization in catalyst design, aiming to optimize multinary metal alloy catalysts for fuel cells [22]. They employed a computational framework combining density functional theory (DFT) calculations, machine learning-driven kinetic modeling, and Bayesian optimization, efficiently identifying optimal catalyst compositions in vast multi-metallic spaces, thereby accelerating catalyst optimization processes. Kazuki and Yuta demonstrate the effective combination of Bayesian optimization and density functional theory calculations to find the optimal binary alloy catalyst for nitrogen activation, a critical step in ammonia synthesis [23]. Their study showcased how Bayesian optimization surpasses random search in efficiency, highlighting the benefits of data science and computational chemistry in accelerating catalyst research, and underscores plans for future exploration of multi-objective optimization for ammonia synthesis.

However, conventional techniques such as OFAT, DoE, and Bayesian optimization cannot efficiently leverage and synthesize the wealth of information embedded in the rapidly expanding body of scientific literature on catalyst synthesis. To bridge the gap mentioned above, many scientists have utilized the power of Artificial Intelligence (AI) to retrieve information from broad corpora of scientific literature. The Large Language Model (LLM) is a type of transformer model that is capable of modeling the probability of a sequence of tokens in texts [24]. By leveraging large-scale data and massive models, it has effectively overcome several long-standing challenges in the field of Natural Language Processing (NLP), such as occasional fluency issues, lack of common knowledge, and unable to remember the context from previous sentences. ChatGPT and its successor, GPT-4, presented by OpenAI, are leading LLMs that are capable of understanding, generating, and translating human language, performing sentiment analysis, and answering questions [25]. Approaching human-level ability across many expert domains, GPT-4 can accomplish complex tasks in chemistry purely from human instructions, potentially setting the stage for transformative advancements in the field of catalyst synthesis. Jablonka et al.'s work demonstrated GPT-3's impressive capability to match conventional machine learning models in chemical property prediction and molecule design tasks, despite using fewer data points, by effectively identifying correlations within textual data [26]. On the other hand, researchers successfully utilized LLMs and particularly developed MolReGPT, which leverages prompts and retrieval methods for translation between molecules and text descriptions, surpassing other models in performance without the need for fine-tuning [27].

By addressing the limitations of traditional chemical space construction and leveraging the strengths of active learning and Bayesian Optimization, the proposed AI workflow to expedite catalyst optimization at the particle scale could prove transformative, offering avenues for swift, resource-efficient, and high-precision catalyst synthesis. This workflow hinges on the convergence of advanced language understanding, Bayesian optimization, and active learning loop methodologies, operating in harmony to ascertain optimal solutions for synthesis parameters that will affect the structure of catalyst particles, consequently affecting their activity. This AI workflow enables the effective integration of knowledge extracted from a wide range of literature with practical experimentation. By fusing the text-understanding capabilities of LLMs, parameter optimization of BO, and the adaptive sampling of an active learning loop, this workflow offers a highly adaptive, robust, and efficient approach to optimizing the synthesis of catalysts.

This article is structured as follows: detailed methodology can be found in Section 2; the results of a demonstrative case study are presented and discussed in Section 3; and conclusions are provided in Section 4.

## 2. Methodology

An illustration of the proposed AI-driven workflow is shown in Figure 1. The backbone of this workflow consists of two key components: the utilization of a large language model for constructing the chemical space and the application of Bayesian active learning for multi-objective optimization within that space. The large language model, ChatGPT trained on an extensive corpus, including scientific literature, is capable of understanding, analyzing, and extracting crucial data and concepts from a vast number of research papers pertinent to catalyst synthesis. This capability enables the AI to build a comprehensive and multi-dimensional search space for catalyst synthesis, encompassing various catalyst types, synthesis methods, reaction conditions, and their corresponding performance metrics. The Bayesian optimization methodology is brought to bear, which uses the information gathered by the large language model to direct the exploration and exploitation of the search space. Bayesian optimization techniques exploit probabilistic models, primarily Gaussian Processes, to approximate the unknown and complex function that maps catalyst synthesis parameters to performance metrics. The AI workflow is then augmented by an active learning loop, which functions to iteratively improve the optimization model. The loop initiates with the Bayesian optimization proposing an optimal set of condition parameters based on the current understanding of the objective function. These suggested parameters are then employed in real-life experiments to synthesize and evaluate catalysts. The performance metrics gathered from these experiments are fed back into the AI model, updating the understanding of the search space and the performance mapping function. Consequently, the loop allows for the AI system to learn dynamically from each experiment, constantly refining its predictions and progressively converging to the global optimum.

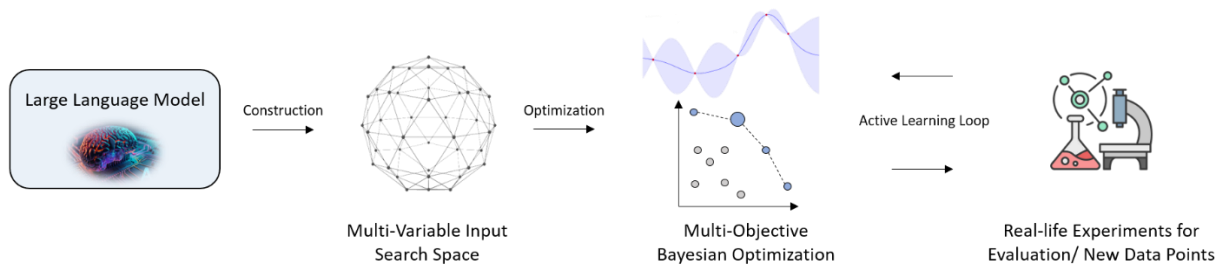

Figure 1. Schematic diagram of the proposed AI-driven workflow for catalyst synthesis optimization.

## 2.1. Large Language Model (LLM) in Chemical Space Construction

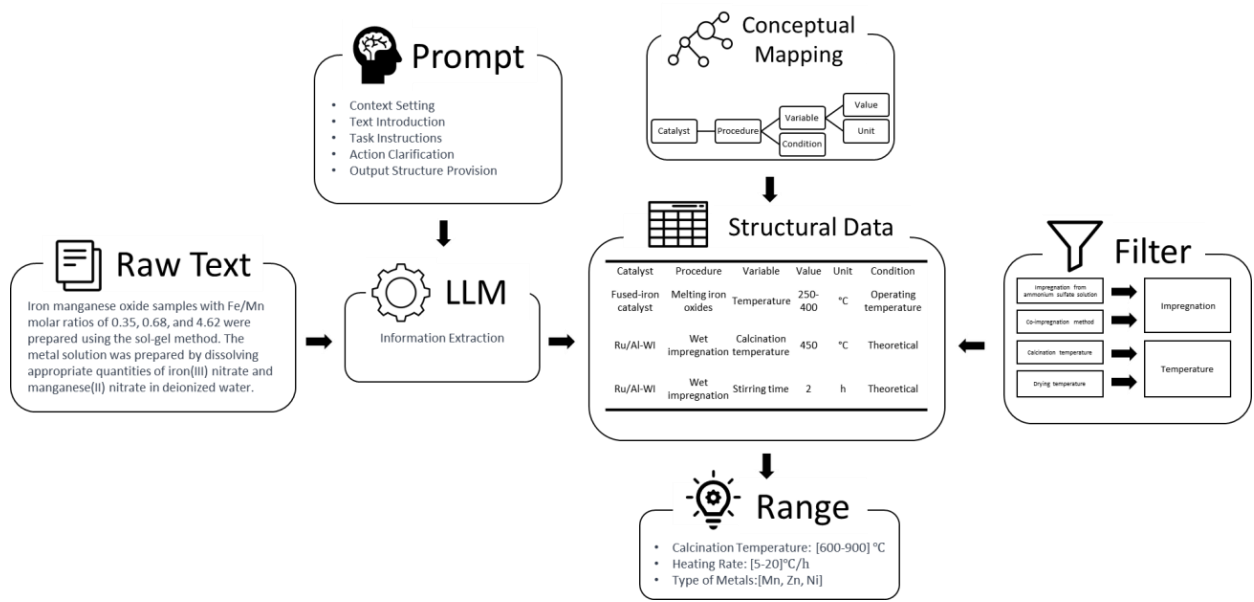

Figure 2. Process flow diagram of information extraction

2.1.1. Collecting Literature Data

Typically, the search space in Bayesian optimization is designed by experts according to their experiences. However, scientific literature contains rich and in-depth information about catalyst preparation and associated variables, which we aim to utilize for the construction of our search space. We place particular emphasis on the diversity of the input data to ensure a comprehensive search space.

The Elsevier Text Mining API was utilized to search the Science Direct database for academic articles relevant to the study. A generalized query string template, TITLE-ABS-KEY('Reaction Type' AND 'catalyst'), was used to capture a diverse range of articles. This template was applied using the demonstrative example of TITLE-ABS-KEY('ammonia synthesis' AND 'catalyst'), which initially yielded 2410 entries. Although the broad search term allows for a comprehensive collection, it may also include articles of varying relevance to the core research questions. The primary criterion for inclusion was the availability of articles in XML format to facilitate more efficient and precise text cleaning and segmentation, reducing the dataset to 603 articles that met this XML criterion. Additionally, while ACS Publications indeed hosts a large number of relevant articles, this study relied solely on Elsevier's Science Direct due to limitations in accessing ACS's text mining API. It should be noted that no manual screening was applied; instead, keyword-based automated filtering identified articles related to catalyst preparation.

After automatically gathering articles on ammonia synthesis catalysts, each literature is segmented into sections that focus on methods and experimental procedures. The filtering process was conducted using a list of keywords to detect catalyst preparation-related phrases, which was designed by sampling the section title from the literature dataset.

2.1.2. Designing Prompts for Data Extraction

Prompts, in the context of LLMs, are structured inputs designed to guide the model in generating specific and desirable outputs. Given that LLMs require guidance to understand the type and structure of the information to be extracted, we created specific prompts to assist in this process. The design of these prompts was influenced by a thorough literature review and expert consultations. Each step in this process was tailored to set a context and provide clear instructions to guide the LLMs, in our case the ChatGPT model.

- Context Setting: You are a researcher in the field of [ammonia catalyst synthesis] who is good at extracting information from text.
- Text Introduction: This is the text which may or may not include the catalyst preparation methods. + <input text>
- Task Instructions: A comprehensive multi-step guide directs the model to analyze the text and extract relevant information, structured depending on the text content.
- Action Clarification: The model's capabilities and tasks are reiterated, clearly defining its role in identifying, extracting, and organizing relevant information.
- Output Structure Provision: A sample table is provided, offering a template for the desired output, guiding the structure and organization of the extracted information.

An example of the prompt used is shown below:

> "You are a researcher in the field of [ammonia catalyst synthesis] who is good at extracting information from text. This is the text which may or may not include the catalyst preparation methods." + <paragraph> +
>
> "Please follow these steps to extract information from the given text:

Start:

1. Analyze the text whether it is about catalyst preparation or not.

2. IF the text is about catalyst preparation THEN proceed to step 3 ELSE go to step 6.

3. Identify the process variables: temperature, pressure, duration of each step, rate of temperature change, reactant proportions, order of reactant addition, stirring speed, pH, concentrations, and additional parameters.

4. Extract these values along with their units.

5. Organize these into a table with the following columns: Catalyst Name, Specific Procedure Step, Process Variable, Numeric Value, and Unit. Then, END.

6. IF the text is not about catalyst preparation, output a blank table.

7. END

You can perform the following actions:

- Identify whether the text is about preparing or synthesizing catalysts.
- Extract related information from the given text and given sample table format.
- Organize related information into the desired format.
- Extract the number and unit from the related information.
- Fill in the table with the desired information.
- Write down the condition of extracted properties in the condition column.
- Output the table.

Sample Table:

| Catalyst Name | Specific Procedure Step/Method | Process Variable | Numeric Value | Unit |
| ZnCrOx and MnZnCrOx | Co-precipitation method | Temperature | 343 | K |
| ZnCrOx and MnZnCrOx | Co-precipitation method | Duration | 30 | min |
| ZnCrOx and MnZnCrOx | Co-precipitation method | pH | 8 | |
| ZnCrOx and MnZnCrOx | Co-precipitation method | Aging | 3 | h |
| ZnCrOx and MnZnCrOx | Co-precipitation method | Calcination temperature | 500 | °C |
| MnZnCrOx | Addition of Mn(NO3)2 aqueous solution | Mass of Mn(NO3)2 | 0.26 | g |
| MnZnCrOx | Addition of Mn(NO3)2 aqueous solution | Concentration of Mn(NO3)2 solution | 50 | wt% |
| ZSM-5 zeolite | Hydrothermal method | Temperature | 180 | °C |
| ZSM-5 zeolite | Hydrothermal method | Duration | 48 | h |
| ZSM-5 zeolite | Hydrothermal method | Calcination temperature | 550 | °C |
| OX-ZEO catalysts | Dual-bed mixing method | Mass ratio of oxide/zeolite | 1 | |
| OX-ZEO catalysts | Granule mixing method | Mass ratio of oxide/zeolite | 1 | |

| OX-ZEO catalysts | Powder mixing method | Mixing duration | 30 | min |
| OX-ZEO catalysts | Powder mixing method | Pellet size | 20-40 | mesh |"

2.1.3. Conceptual Mapping of Extracted Information

Recognizing the critical role of variable relationships in catalyst preparation, we organized the extracted data into a conceptual map. This map was created using Neo4j, a graph database management system, and parameters were selected based on frequency analysis and expert opinion [28]. The map effectively illustrates relationships between catalyst types, preparation procedures, procedure variables, numeric values, units, and applied conditions, as shown in Figure 2.

2.1.4. Quality Control

At present, the Large Language Models (LLMs) are in the initial stages of fully understanding the context of catalyst preparation procedures and associated variables. To ensure the reliability and applicability of the information generated by ChatGPT, domain experts meticulously review extracted parameters, such as Activation Temperature (AT) and Activation Duration (AD). Their expertise aids in determining the range and feasibility of these parameters in real-world chemical processes, thereby aligning the AI-extracted data with current scientific understanding. To enhance the robustness of the information extraction process, two distinct filters have been designed to systematically categorize procedures and variables into their relevant categories. For a more concrete understanding, the following examples are provided.

For the process of melting iron oxides with structural promoters such as AlO, and CaO, and an activating promoter like KO, the specific procedure is classified as 'Combustion'. Similarly, in the 'Wet impregnation method', the specific procedure is 'Impregnation', and the variables such as 'Calcination temperature' and 'Drying temperature' are classified as 'Temperature', while 'Stirring time' is classified as 'Duration'.

2.1.5. Vectorization for Optimization

Given the need for a mathematical representation of our data to input into Bayesian optimization models, we converted our structural data into a multi-dimensional vector space. The min-max normalization method was used for continuous variables to define the range of space. For discrete variables, we included all options mentioned in the literature. This strategy facilitates an exhaustive representation of potential solutions within the "chemical space" associated with our catalyst.

## 2.2. Bayesian Active Learning

Active learning, a crucial aspect of machine learning, is particularly instrumental in the area of optimal experimental design [29, 30]. The proposed Bayesian optimization loop is shown in Figure 3. This systematic approach helps identify the most beneficial experiments to be conducted next based on predefined objectives. One technique that has gained considerable attention recently is Bayesian Optimization (BO), an active learning technique that guides experimentalists in the lab to optimize unknown functions [31]. BO balances the exploration of the unknown function with experiments that exploit prior knowledge to pinpoint extrema. They operate with minimal data and use heuristic-based searches for the most informative observations [32]. This approach is versatile and can be applied to diverse search spaces, including arbitrary parameterized reaction domains, making it well-suited for optimizing the variables of catalyst synthesis. The strength of BO lies in its capacity to handle noise, high dimensionality, and the nonlinearity of the objective function, making it highly suitable for optimizing complex processes such as catalyst synthesis.

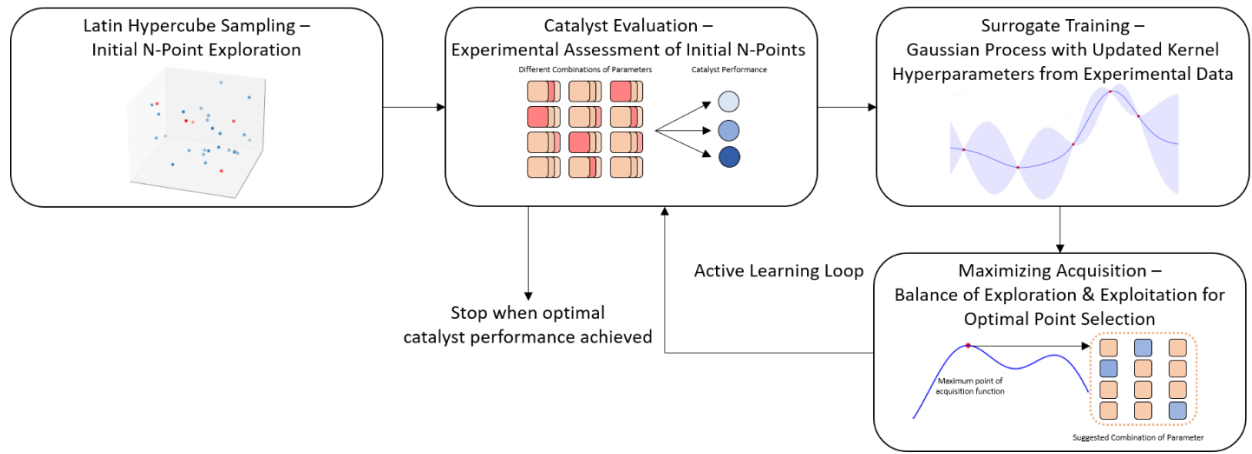

Figure 3. Detailed depiction of the Bayesian optimization loop utilized within the proposed AI workflow.

2.2.1. Initial Sampling Strategies

In the realm of Bayesian Optimization, the selection of initial sampling points plays a pivotal role in determining the overall effectiveness and efficiency of the optimization process [30]. These initial points from the search space form the foundation for the subsequent iterative learning and optimization steps, essentially guiding the model's exploration of the search space. This leads to the importance of employing robust initial sampling strategies, as they can significantly influence the optimization outcome.

Among the various initial sampling strategies available, Latin Hypercube Sampling (LHS) [33] has been utilized in this work. LHS, a stratified sampling technique, stands out due to its capability to generate a well-distributed set of initial points across the entire search space. Unlike other methods that randomly select initial points, LHS ensures that the sampled points span the search space as evenly as possible, which ensures a balanced exploration of the search space right from the outset, thus increasing the probability of identifying the optimal solution. Given the vastness and the high-dimensional nature of the catalyst parameter space, a well-distributed initial sampling, as afforded by LHS, ensures a broad and diverse representation of the search space [34]. This is especially advantageous when dealing with complex, multi-step catalyst synthesis processes that involve numerous interrelated parameters.

2.2.1. Surrogate Model of BO

The implementation of a surrogate model is a fundamental facet of BO, with the Gaussian Process (GP) model adopted in this research because of its proficiency in managing uncertainty and the multifaceted nature inherent to catalyst synthesis optimization [35].

$$\begin{bmatrix} f(x_1) \\ \vdots \\ f(x_n) \end{bmatrix} \sim N\left( \begin{bmatrix} m(x_1) \\ \vdots \\ m(x_n) \end{bmatrix}, \begin{bmatrix} k(x_1, x_1) & \cdots & k(x_1, x_n) \\ \vdots & \ddots & \vdots \\ k(x_n, x_1) & \cdots & k(x_n, x_n) \end{bmatrix} \right) \quad (1)$$

A GP surrogate model offers a non-parametric, probabilistic technique to model the elusive function connecting the parameters of catalyst synthesis to performance metrics[21]. GP is formally defined as eq (1), where $X = \{x_1, \ldots, x_n\}$ is the vector of process variables. In this formulation, m(X) represents the mean function, encapsulating the expected value of GP across all elements of X. $k(X, X')$ is the covariance matrix, providing a measure of interrelation for all possible pairs within the set of process variables $(X, X')$. The GP model's distinct advantage is its capability to estimate prediction-associated uncertainty, making it ideal for Bayesian Optimization that intrinsically requires balancing exploration and exploitation. This quantification of uncertainty enables the intelligent selection of subsequent data points, harmonizing the necessity for further exploration with optimization goals. The choice of mean function and covariance (or kernel) function in the GP model is instrumental in shaping the model's properties and subsequent performance. In this study, a constant mean function was selected for our Gaussian Process surrogate model, premising that the average value across the entire search space remains constant. This assumption, while seemingly simple, is often a practical and effective starting point, particularly when undertaking new catalyst development projects. In such scenarios, the prior knowledge about the catalyst's performance landscape is typically sparse or even absent, making it challenging to substantiate more complex assumptions about the mean function. Therefore, a constant mean function provides a reasonable and unbiased baseline from which to start our optimization process and allows the model to learn and update its understanding of the objective function as more data is gathered.

$$k(X, X') = \sigma^2 \left(1 + \frac{\sqrt{5}r}{\ell} + \frac{5r^2}{5\ell^2}\right) \exp\left(-\frac{\sqrt{5}r}{\ell}\right) \qquad (2)$$

Additionally, for the covariance function, a comparative assessment was carried out among a selection of Matern functions characterized by differing degrees of smoothness, denoted by $\nu$. The Matern 5/2 kernel, expressed in eq 2, where $r$ is the Euclidean distance between the points X and $X'$, $r = \|X - X'\|_2$, $\sigma^2$ is the signal variance and $\ell$ is the length scale. Kernel emerged as the optimal choice for this particular catalyst synthesis optimization issue, striking an effective balance between model adaptability and computational efficiency. The Matern 5/2 kernel, with its intermediate level of smoothness, tends to perform well in practice, especially when handling noisy data or navigating high-dimensional search spaces. The parameter $\nu = 0.5, 1.5, 2.5$ determines the smoothness of the function. The $\nu = 0.5$ case corresponds to an absolute exponential (Laplace) kernel, representing a very non-smooth process. On the other end, infinity corresponds to the Radial Basis Function (RBF) kernel, representing an infinitely smooth process. The other two cases ($\nu = 1.5$ and $\nu = 2.5$) offer a balance between smoothness and flexibility, with the Matérn $\nu$

= 2.5 kernel being particularly highlighted due to its robustness in handling noisy data and high-dimensional spaces.

Moreover, GP predominantly operates on continuous variables, necessitating the conversion of categorical variables into a continuous format [36, 37]. During the initial stages of optimization catalyst synthesis, the volume of experimental data is typically limited, especially in the field of the development of new catalysts. In these low-data regimes, simpler representation methods such as one-hot encodings (OHE) often yield significant results, or even surpass, the performance of more complex and resource-intensive descriptors [38, 39]. The handling of categorical variables which are identified by the first step of the AI workflow as some of the important variables, using the method of one-hot encoding adds to the flexibility of the GP model, broadening its applicability to a wider range of scenarios. For example, in the synthesis of catalysts, where various solvents and metals are used, these categorical variables can be suitably represented in binary form via one-hot encoding. This binary representation designates the presence or absence of specific solvents and metals. Beyond a binary representation, one-hot encoding can also accommodate the representation of multiple types of solvents and metals. Therefore, it does not just provide binary inclusion-exclusion information but extends to symbolize a multitude of categories within each variable. This way, the use of one-hot encoding contributes to an increasingly flexible Gaussian Process model, accommodating a broader span of input scenarios and enhancing its performance within the complex and multifaceted landscape of catalyst synthesis. This incorporation further bolsters the robustness of our GP-based Bayesian optimization framework.

Hence, the integration of the Gaussian Process surrogate model, a constant mean function, the Matern 5/2 covariance function, and effective handling of categorical inputs generate a robust, adaptable, and potent framework to model the intricate, high-dimensional, and uncertain search space intrinsic to catalyst synthesis. This firm foundation enhances the effectiveness of subsequent optimization stages, thus improving the likelihood of converging towards the global optimum.

2.2.2. Acquisition Function of MOBO

Bayesian optimization methodology centers around the acquisition function, a key component that determines the next set of experimental parameters to investigate. For single-objective optimization, the Expected Improvement (EI) acquisition function is often used due to its ability to balance exploration, targeting areas with significant predictive uncertainty, and exploitation, focusing on regions predicted to offer high performance [40]. This balance ensures effective and systematic use of resources. However, catalyst synthesis represents a complex, multi-objective optimization task, which is influenced by a variety of performance factors such as catalyst activity, selectivity, stability, and adaptability under various reaction conditions. The relationship between these often conflicting parameters is well-captured by the concept of Pareto optimality, which represents a set of non-dominated solutions where improvement in one objective invariably affects others [41]. By aiming to find solutions on or close to the Pareto front, a set of optimal solutions can be found where a decision-maker can choose depending on their specific preferences or requirements [40]. This provides a more comprehensive view of the possible optimal solutions, rather than a single "best" solution, which might not exist in multi-objective problems due to the conflicting nature of the objectives. To elaborate, activity and stability are the two parameters to be optimized. Any solution that increases the activity without decreasing the stability, or increases the stability without decreasing the activity, is considered Pareto optimal.

When dealing with multi-objective optimization, a more nuanced approach to the acquisition function is needed. This approach must balance multiple objectives, either by assigning specific weights to the objectives or by applying the principle of Pareto optimality. This is where the Expected Hypervolume Improvement (EHVI), a multi-objective acquisition function, comes into play [42]. The EHVI function considers the entire Pareto front, providing a comprehensive view of all optimal solutions. It is similar to the improvement in the EI acquisition function used for single-objective optimization, the difference is that in EHVI, the improvement refers to the expected increase in the dominated hypervolume if a new sampling point were to be incorporated [43, 44]. This means that EHVI considers not just the mean and variance predictions of the underlying Gaussian Process model, but also the correlation between multiple objectives. It guides the exploration through the high-dimensional parameter space to find the next promising point for exploration.

2.2.3 Active Learning Loop

In the Bayesian optimization framework applied to catalyst synthesis parameter optimization, the active learning loop is the crucial phase that bridges the gap between theoretical predictions and practical implementation [44]. Here, the selected set of parameters provided by the acquisition function is translated into a tangible experimental setup. Initially, experiments are conducted using the suggested combination of parameters. This step embodies the process of exploration and exploitation as indicated by the Gaussian Process surrogate model's predictive framework, which provides statistical estimates based on accumulated knowledge. This stage entails meticulous catalyst synthesis according to stipulated parameters and conditions, offering practical validation for the theoretical predictions of the model.

Once the catalyst is synthesized, its performance is subjected to a comprehensive evaluation. Various performance metrics, such as catalyst activity, selectivity, stability, and adaptability under diverse reaction conditions, are measured. The precision in determining these metrics is of significant importance as they are incorporated back into the model, thereby directly influencing future predictions and optimization steps. Following the catalyst performance evaluation, the GP surrogate model is updated with the newly obtained experimental data. By accommodating this fresh information, the model's capability to predict the performance of different parameters is continually refined. A defining aspect of the active learning loop is the decision of when to terminate the process. The loop can conclude when the optimization process reaches a predetermined threshold of catalyst performance or a specific level of improvement. Alternatively, the loop might cease when the maximum allowable number of experiments is reached. The convergence criteria, while generally pre-established, should also incorporate practical factors such as resource constraints and the urgency of achieving the desired performance.

The active learning loop thus forms the backbone of the Bayesian optimization methodology. It fosters a systematic and efficient trajectory through the complex, high-dimensional parameter space of the catalyst synthesis process. By facilitating a robust cycle of prediction, experimentation, and learning, it paves the way for optimizing catalyst performance, offering significant potential to enhance the effectiveness of catalyst synthesis.

## 3. Result and Discussion

In this section, we explore the practical implementation and implications of the delineated AI Workflow for Catalyst Synthesis. We demonstrate the method's effectiveness and potential in new catalyst design through an Ammonia Synthesis case study.

The rising clean energy trend has spurred intense research into green hydrogen generation and its conversion into green ammonia. The drive towards decentralized green ammonia production necessitates equipment miniaturization and a reduction in reaction temperature and pressure This need aligns with contemporary catalyst requirements that prioritize milder operating conditions, robustness to potential catalyst poisoning from trace water oxygen in green hydrogen, and cost-effectiveness. Notably, nitride catalysts are emerging as significant players in ammonia synthesis. For instance, efforts have been made to construct a Ni/LaN system, leveraging the nitrogen vacancies of LaN to weaken N-N triple bonds, promoting $N_2$ dissociation assisted by Ni's facilitation of $H_2$ dissociation, a synergic approach albeit hindered by Ni's weak interaction with $N_2$, restricting further enhancement of the catalytic activity of the system [45]. Besides, the $Co_3Mo_3N$ catalyst stands out, exhibiting promising attributes aided by Cs or K promoters, nearing the performance of commercial Fe-based catalysts, although grappling with cost impediments and gaps in high-pressure activity data and resilience against water and oxygen[46]. In this study, we focused on the synthesis of the CoMo bimetallic nitride catalyst known for its higher ammonia synthesis activity[47]. This initiative involves the optimization of the activation step parameters to find an optimal set, promising a rich yield of ammonia. The process, characterized by a multitude of phases contributing to the catalyst's activity, requires parameter tuning to find the optimal parameter combination. Addressing these high-dimensional tasks necessitates the utilization of data-driven approaches and advanced machine-learning models. In this regard, the AI Workflow outlined in this study could be instrumental in overcoming these challenges. Such a method has the potential to streamline the design and synthesis of new catalysts, thereby improving the efficiency and reducing the time and cost associated with catalyst development.

3.1 Catalyst Preparation

The genesis of the $Co_3Mo_3N$ (CMN) catalyst, central to ammonia production, begins with the preparation of CoMo bimetallic oxide, $CoMoO_4$ precursor through the hydrothermal method. We find the Co/Mo catalyst preparation process at a feed ratio of 1:1 to be a fitting example. Precisely weigh 5.8026 g of $Co(NO_3)_2 \cdot 6H_2O$ and 4.8390 g of $NaMoO_4 \cdot 2H_2O$. Each is dissolved separately in 60 mL of deionized water, then stirred for 5 minutes at a speed of 500 rpm. Following this, the two solutions are combined and stirred continuously for an additional 10 minutes at the same speed. The mixed solution is then transferred into two 100 mL hydrothermal kettles, each containing approximately 60 mL of the solution. Place the hydrothermal kettles in a blast drying oven to undergo a hydrothermal reaction at 160 °C for 6 hours. Remove the kettles once they have completely cooled. The precipitate slurry obtained from the hydrothermal kettles is subjected to solid-liquid separation using a centrifuge operating at 5000 rpm for 5 minutes. Following centrifugation, discard the supernatant and add a certain amount of deionized water to mix evenly with the precipitate at the bottom of the centrifuge tube. Disperse ultrasonically for 5 minutes to thoroughly wash the precipitate before centrifuging again. Repeat this washing step three times. Transfer the thoroughly washed precipitate into a blast drying oven and dry for 12 hours at a set temperature of 100 °C. After drying, grind the violet blocky solid into a powder. Transfer this powder into a crucible for calcination in a muffle furnace at 450 °C for 6 hours, with a heating rate of 10 °C/min. After calcination, allow it to cool naturally to room temperature in the open air to obtain a light violet catalyst powder. Transfer the solid from the crucible to a sample bottle, marking the completion of the catalyst precursor preparation.

Figure 4 showcases the multi-stage synthesis process of the CoMo bimetallic nitride catalyst. Each phase, marked by distinct parameters illustrated in the figure, holds a role in determining the overall catalyst performance, albeit not equally. It should be noted that while there are other parameters influencing the synthesis process, including those at the hydrothermal stage, we strategically elected to focus on the activation step because modifying parameters in the hydrothermal stage generally necessitates a substantially extended timeframe to observe the resultant activity alterations, from several days to weeks. Consequently, it induces an exponential increase in both time and material costs. the activation step is where nitride is incorporated into the catalyst, a process in determining the structure of the resultant catalyst. The parameters chosen for optimization in our study are intrinsically linked to the control of this nitride introduction process, thereby holding a direct influence over the catalyst's final structure.

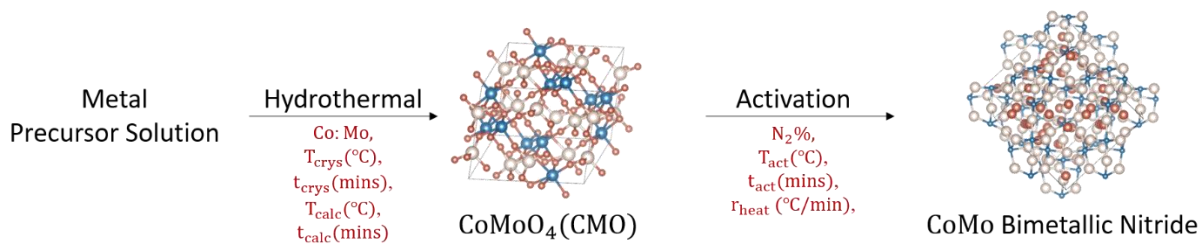

Figure 4. Schematic diagram illustrating the multi-stage $Co_3Mo_3N$(CMN) catalyst synthesis process for ammonia production.

3.2 Data Extraction for Search Space Construction

Our study encompasses a diverse dataset by extracting information from literature using ChatGPT, that accounts for 774 distinct catalysts and 737 unique procedure steps, offering a comprehensive overview of various catalysts and procedures involved in ammonia synthesis processes. Table 1 and Table 2 serve as an illustrative summary of the dataset.

As shown in Table 1, the analysis of catalyst frequencies reveals that "Ammonia Catalyst" appears to be the most prevalent, finding use in 437 instances. This is succeeded by "Ru/MgO" and "Ammonia Synthesis Catalyst", appearing in 47 and 43 instances, respectively. Despite the considerable variety in the dataset, certain catalysts emerge as common across processes, with these three showing the highest frequency of usage.

Similarly, Table 2 outlines the frequency of different procedural steps. The "Impregnation method" stands as the most common, with 240 instances. This is followed by the "Synthesis" (187 instances), and "Reduction" (104 instances) methods, underscoring their significance in the analyzed chemical processes. Among the less frequent procedures, "Fixed-bed flow reactor" and "Synthesis with aqueous solutions of MnSO·HO and KMnO" appear in 24 instances each.

Table 1 Frequency of Catalyst Names Extracted by ChatGPT from Literature

| Catalyst Name | Counts |
|---|---|
| Ammonia Catalyst | 437 |
| Ru/MgO | 47 |
| Ammonia Synthesis Catalyst | 43 |
| CoMoN | 37 |
| Fe-containing catalysts | 37 |
| Ru/CeO | 35 |
| NH synthesis catalyst | 33 |
| Iron catalyst | 32 |
| Graphitic-nanofilaments | 30 |
| Cs-Ru/CeO | 29 |
| NiMoO | 27 |
| Ruthenium catalysts | 27 |
| Ru/C | 26 |
| CeO | 23 |
| N-doped carbon support | 22 |
| CdS/CNS | 20 |
| Cs-Ru/BaTaO | 20 |
| FT Catalysts | 19 |
| BCDs/NiCoO/CC | 18 |
| Cu-NC | 18 |
| Plasma Synthesis of Ammonia | 18 |
| TiOH/K/Ru | 18 |
| Co/SiO and Co/SiO-AlO | 17 |
| CoMoOÂ·HO | 17 |

Table 2 Frequency of Procedure Steps Extracted by ChatGPT from Literature

| Procedure Step | Counts |
|---|---|
| Impregnation method | 240 |
| Synthesis | 187 |
| Reduction | 104 |
| Hydrothermal method | 91 |
| Catalyst Preparation | 82 |
| Co-precipitation method | 71 |
| Ammonia Synthesis | 66 |
| NH synthesis | 55 |
| Solvothermal method | 52 |
| Calcination | 48 |
| Preparation | 46 |
| Impregnation | 43 |
| Drying | 42 |
| Ammonia synthesis | 39 |
| Wet impregnation method | 39 |
| Reduction in hydrogen | 33 |
| Incipient wetness method | 32 |
| Reduction method | 32 |
| Incipient wetness impregnation method | 29 |
| Incipient wetness impregnation | 27 |
| Precursor preparation | 27 |
| Nanoreplication method | 25 |
| Fixed-bed flow reactor | 24 |
| Synthesis with aqueous solutions of MnSOÂ·HO and KMnO | 24 |

Our data analysis from 603 articles on ammonia synthesis catalysts yielded significant insights into the multidimensional parameters shaping catalyst synthesis. These parameters included

Activation Pressure (AP), Activation Duration (AD), Activation Temperature (AT), and Heating Rate (HR).

The optimization procedure targets key variables within this reduction step: activation pressure, activation temperature, heating rate, and activation duration, which emerged as dominant forces shaping the catalyst's structure and consequently, its performance in ammonia synthesis. These variables were identified as impactful parameters through the application of the large language model ChatGPT, which intelligently parses through extensive literature to discern the most influential factors in the catalyst synthesis process.

We observed that the Activation Pressure ranged from 3 to 10 MPa, which underscores the importance of accurate pressure control for optimal catalyst synthesis. Activation Duration, another crucial variable, presented a broader range from 0.5 to 8 hours, reflecting a wide temporal span for catalyst activation. The parameter of Activation Temperature showcased substantial variation, ranging from 200°C to 700°C. Meanwhile, the presence of lower temperatures suggests that certain catalysts can be synthesized and activated under relatively mild conditions. The Heating Rate, presented in °C/min, had a narrower range of values, primarily between 1 and 10.

Due to the limited amount of literature collected, the results extracted do not fully represent the entire search space. Enlarging the search space in terms of Activation Temperature is conceptualized with a foresight that higher temperatures could foster the formation of different phase structures, which in turn might influence the catalytic activity positively. This exploration promises the discovery of potentially favorable conditions that were not contemplated within the original bounds, possibly leading to groundbreaking insights into the activation parameters and their roles in synthesizing efficient catalysts. On a parallel note, initial rounds of experimentation suggested potential improvements in catalytic performance at different Heating Rates However, initial rounds of experimentation suggested potential improvements in catalytic performance at higher Activation temperatures and different Heating Rates. Therefore, in our effort to explore a larger space, we have appropriately expanded the search parameters. Specifically, we have increased the Activation Temperature up to 900°C and the Heating Rate up to 20 °C/min. This expansion will hopefully provide a broader understanding of catalyst activation, beyond the constraints of our initial findings.

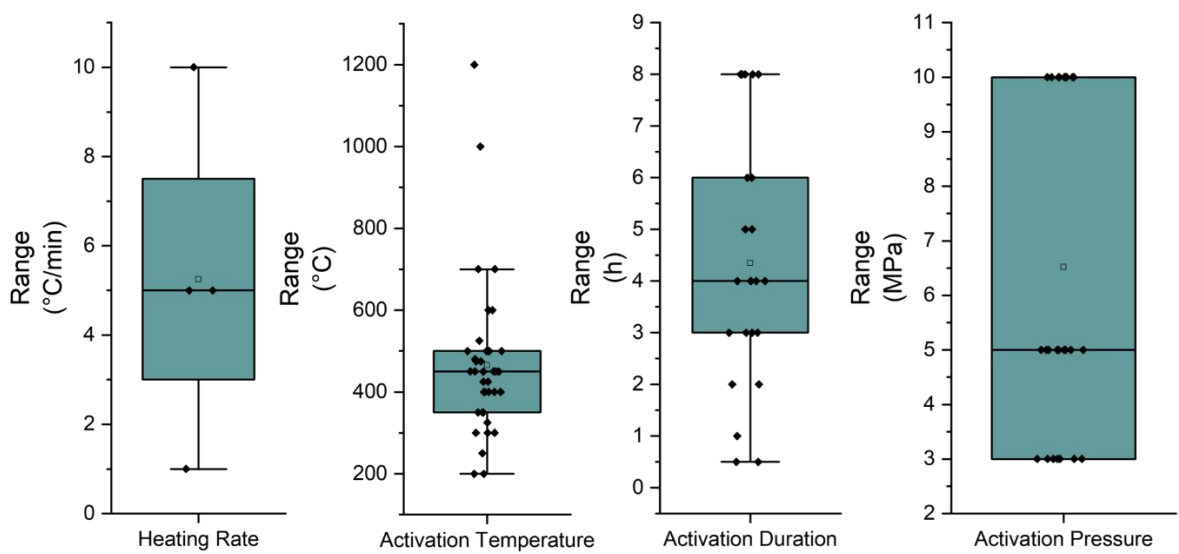

Figure 5. Statistical analysis of catalyst synthesis parameters collected from the literature

3.3 Performance Analysis: Bayesian Optimization in Active Learning Context

The initialization of data points within the Bayesian Optimization process is visualized in Figure 6, specifically focusing on the pairwise distribution of four critical parameters: Activation Pressure (MPa), Activation Duration (mins), Heating Rate (°C/min), and Activation Temperature (°C). Two different sampling strategies have been employed: Random Sampling (represented in Teal) and Latin Hypercube Sampling (LHS, represented in Blue). The upper and lower triangles of the grid contain scatter plots illustrating the relationships between each pair of parameters for each sampling method. Along the diagonal, histograms provide a clear overview of the frequency distribution of each parameter for both sampling strategies.

From this graphical representation, it is apparent that LHS ensures a more uniformly dispersed coverage across the parameter space in contrast to Random Sampling. This superior attribute of LHS for the initialization step effectively minimizes the clustering of sample points, reducing the potential for neglecting important regions of the parameter space. By generating a comprehensive initial exploration of the space, we set a robust foundation for the subsequent steps in the Bayesian Optimization process, enhancing its efficacy and efficiency in locating the optimum. The number of initial points (7 data points here) is determined based on the experimental capacity. Researchers can choose the number of points that align with the number of experiments that can be conducted concurrently or sequentially, thereby effectively utilizing their resources and time.

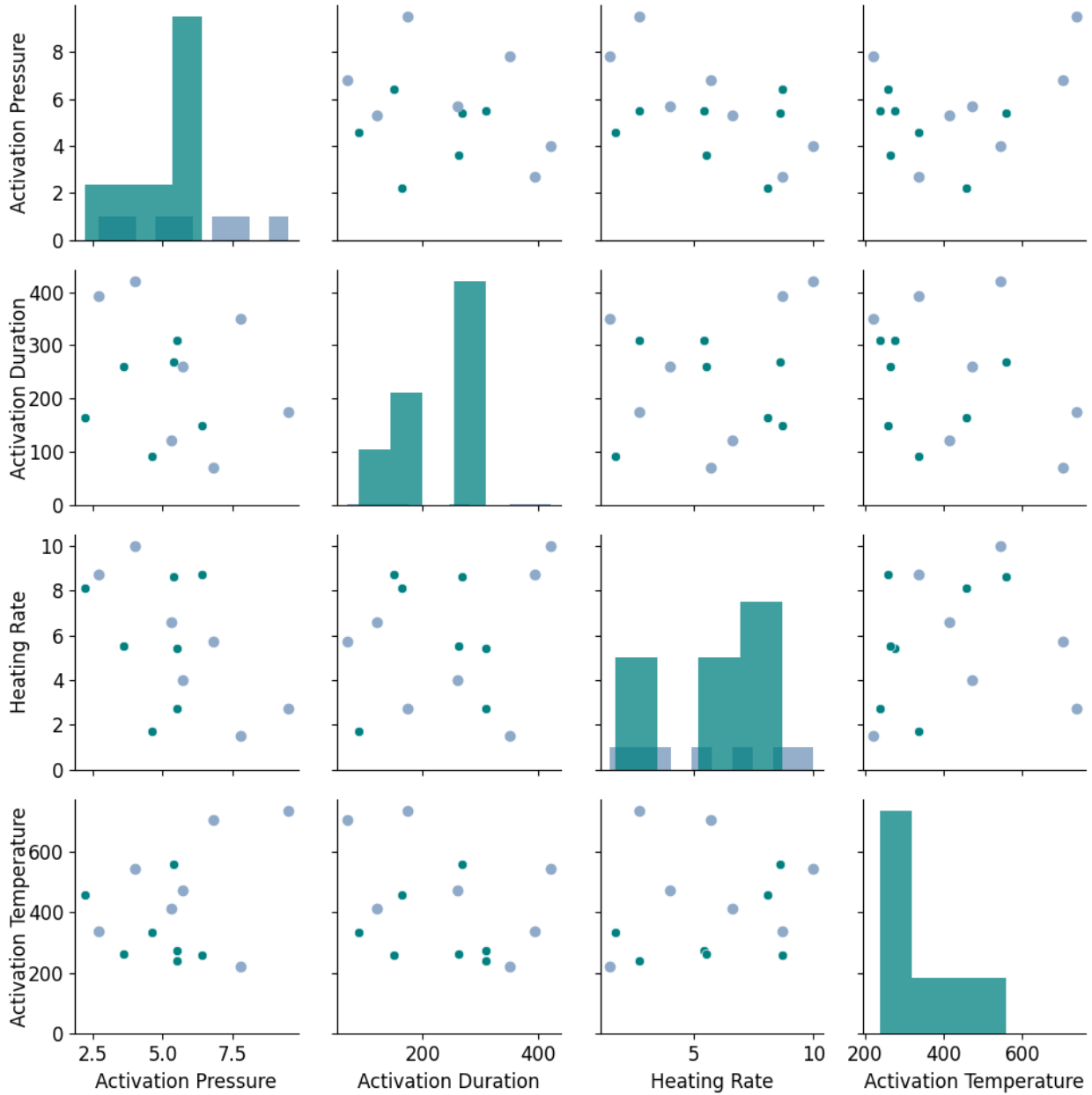

Figure 6. Pairwise distribution of four critical parameters: Activation Pressure (MPa), Activation Duration (mins), Heating Rate (°C/min), and Activation Temperature (°C) visualized through two sampling strategies: Random Sampling (RS, teal) and Latin Hypercube Sampling (LHS, blue).

Following the initialization, researchers embark on the task of evaluating these seven points. Each point corresponds to a unique combination of Activation Pressure, Activation Duration, Heating Rate, and Activation Temperature, which are the variables or parameters under consideration. The researchers conduct experiments under these conditions to evaluate the performance of the system, specifically, the concentration of $NH_3$ in ppm. To maintain uniform characteristics across all trials, we standardized the preparation of the catalyst precursor under a fixed set of conditions. The variables manipulated in each iteration were confined to activation parameters. These were directed by Bayesian optimization. This approach ensured the consistency of the catalyst base, using new samples from the same precursor batch for each experiment, thereby removing any variables introduced by different catalyst types or preparation technologies. The focal point of this optimization was the activation step, pivotal in nitride incorporation into the catalyst and determinant in structuring the resultant catalyst. Thus, our parameter choices for optimization were centrally concerned with modulating the nitride introduction process, directly influencing the final catalyst architecture.

Upon completion of the evaluation, the results are fed back into the Bayesian optimization algorithm. This iterative process of 'suggest-evaluate-update' is the core of active learning in Bayesian optimization. While in a traditional active learning setup, the experiment would be conducted iteratively, with feedback provided to the algorithm after each experiment, in this case, all seven experiments are conducted concurrently. The results of these experiments are then gathered and fed back into the Bayesian optimization algorithm all at once. Even though the algorithm does not get an opportunity to update the surrogate model incrementally after each experiment, it can still gain valuable information from the batch of results to suggest the next most promising point. The advantage of this approach is that it allows for parallelization of experiments, potentially saving considerable time. Furthermore, the convergence of the surrogate model can still be tracked. The surrogate model, which approximates the true but unknown objective function, iteratively becomes more accurate as more data points are evaluated. Its convergence is an essential aspect of Bayesian optimization, as it indicates the algorithm's progress in learning the objective function's landscape. It is important to note that the surrogate model's convergence does not necessarily imply that we have found the global optimum, but it does show that the algorithm is effectively learning from the evaluated data points to suggest better and more informed future points for evaluation. In each iteration, the algorithm suggests the next point (or set of points) to evaluate based on the updated model and an acquisition function. In this case, we utilize Expected Improvement (EI) as our acquisition function. The EI function guides the search towards regions of the parameter space that are expected to offer the most significant improvement over the current best-known performance.

This is a single-objective optimization problem, to maximize the concentration of $NH_3$. Mathematically, our objective function is centered around maximizing $NH_3$ concentration by tuning the activation parameters (Activation Pressure - P, Activation Duration - D, Heating Rate - R, and Activation Temperature - T), and this optimization is constrained within viable and realistic bounds for each parameter, represented as eq (3):

$$\underset{P,D,R,T}{\text{maximize}} f(P, D, R, T) = NH_3 \text{ Concentration} \qquad (3)$$

$$\text{subject to } 2\text{MPa} \leq P \leq 10\text{MPa}$$

$$0.5\text{hours} \leq D \leq 8\text{hours}$$

$$1\,°C/\text{min} \leq R \leq 20\,°C/\text{min}$$

$$200°C \leq T \leq 900°C$$

The algorithm works towards finding the set of parameters that yield the highest concentration, effectively navigating the balance between the exploitation of areas of the parameter space known to yield good results and the exploration of less known areas that may offer further improvements. The final result of this process is an optimized set of conditions that yield the maximum concentration of $NH_3$, given the constraints of the problem.

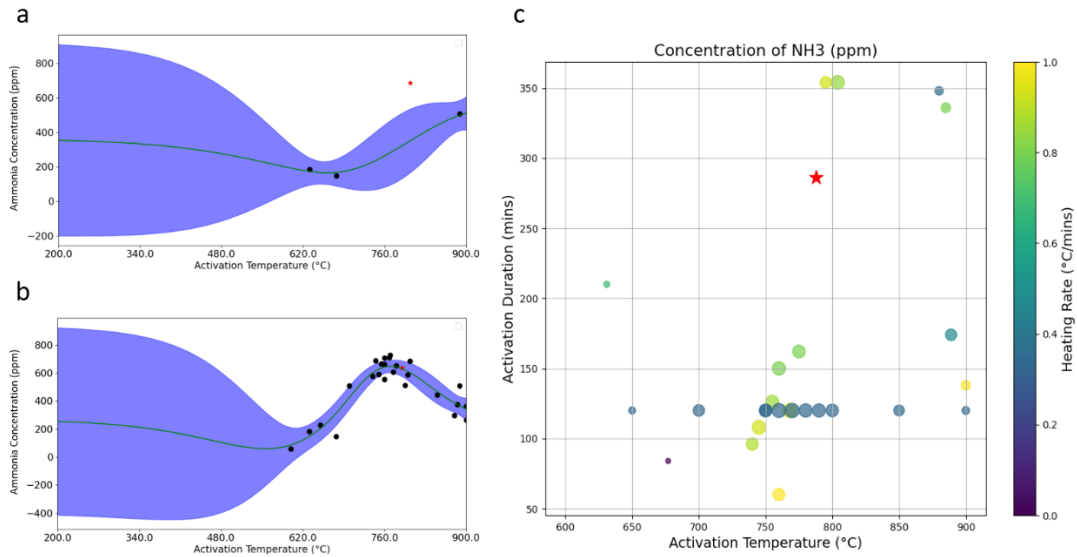

Figure 7. a. The Surrogate model's initial state incorporates just a few data points. b. Transitions to a later stage in the optimization process, where the model has been refined by a significantly larger set of data points. c. Suggested optimal point as indicated by the Bayesian optimization algorithm, along with the set of experimental data points obtained thus far.

Despite this difference in approach, the algorithm continues to accurately track the convergence of the surrogate model, an essential aspect in Bayesian optimization. This model, an approximation of the true but unknown objective function, becomes progressively more accurate as more data points are evaluated. The surrogate model's convergence does not necessarily indicate the discovery of the global optimum but highlights the algorithm's ability to learn effectively from evaluated data points to suggest improved future points for evaluation. In each iteration, the algorithm proposes the next point for evaluation based on the updated model and the acquisition function. As shown in Figure 7a, in the initial iteration, the surrogate model provides a basic estimation of the system's behavior. Given the sparse data, the model is limited in its ability to capture the complexity of the parameter interactions. Despite this limitation, the GP model offers a good starting point for understanding the system's dynamics, focusing on areas that the model perceives as having the most potential for improvement based on the available data. As we move to a later iteration that incorporates a larger number of data points (Figure 7b), the surrogate model becomes significantly more refined. With this additional data, the model is better equipped to capture the system's behavior accurately and construct a more intricate understanding of the underlying relationships between parameters. Consequently, the surrogate model gains increased confidence in its predictions, as indicated by a decrease in the uncertainty around the estimated output.

Figure 7c provides a multi-dimensional representation of the parametric space of the ammonia production process. This visual representation elucidates the correlations between various operating conditions—namely activation temperature, activation duration, and heating rate—and the resultant ammonia concentration. The colors indicate the heating rate of the catalyst, while the size of each data point is proportional to the concentration of $NH_3$ produced under those conditions. The star symbol in red represents the suggested optimal point proposed by the Bayesian optimization algorithm. The proposed optimal point, as determined by the Bayesian optimization algorithm, is graphically depicted within the explored parameter space. This point, characterized by an activation temperature of 787.782 ℃, an activation duration of 286.235 minutes, a heating rate of 14.239 ℃/min,, and an activation pressure of 5.002, is predicted by the surrogate model to yield an ammonia concentration of 730.43 ppm. Importantly, while this point is indeed suggested based on the algorithm's current understanding of the chemical space, its effectiveness, and true performance remain to be validated through further experimental testing.

## 4. Conclusion

In conclusion, the AI workflow proposed for catalyst synthesis optimization represents a compelling intersection of advanced language understanding, Bayesian optimization, and an active learning loop. ChatGPT, as a representative Large language model, has been used to construct a comprehensive and multidimensional search space. We then employed Bayesian optimization for effective multi-objective optimization within this space. Together, these methods demonstrated the potential to streamline the design and synthesis process for new catalysts.

In practical terms, this study has revealed substantial insights into catalyst synthesis for ammonia production – a burgeoning area of clean energy research. Through our analysis of 603 articles on ammonia synthesis catalysts, we have illuminated the significant role of factors such as Activation Pressure, Activation Duration, Activation Temperature, and Heating Rate in the synthesis process. We have also showcased the expansive nature of the catalyst and procedural steps used in ammonia synthesis, reflecting the potential of our methodology to accommodate a diverse range of catalysts and synthesis processes. The practical implications of our work extend far beyond the mere understanding of catalyst design and synthesis. Our results have given us unique insights into the optimization of the ammonia production process. The Bayesian optimization algorithm proposed a set of conditions that, according to the GP surrogate model, and EI as an acquisition function, could result in an enhanced ammonia concentration. Specifically, the model suggested a combination of the parameters, forecasting an ammonia concentration of 730.43 ppm. This prediction, while yet to be verified experimentally, underscores the potential of AI-driven methods in guiding future experimental design.

While this study marks a considerable advancement in utilizing AI for catalyst synthesis, it is imperative to acknowledge the potential challenges and limitations encountered in the current framework. Future work should look at the reliability of published data, which forms the foundation of the LLM model. Any inconsistencies in data could potentially affect the output. Therefore, validations should be made beyond the domain knowledge of the chemist. In addition, there is a clear necessity to expand the AI workflow to deal with multi-objective optimization problems that are frequently encountered in catalyst synthesis. This expansion should also include the ability to handle categorical input parameters, a feature critical for navigating the complexity of chemical spaces. Nevertheless, the AI workflow showcased in this study is a substantial leap forward in the quest for streamlined catalyst design. By effectively distilling insights from an extensive corpus of literature, this workflow demonstrates the potential of AI to supplement, if not replace, the domain knowledge traditionally required in this field. As we venture further into the era of clean energy, AI stands poised to play an increasingly prominent role in catalyzing advancements in catalyst synthesis and beyond.

Looking ahead, our goal is to evolve this AI workflow into an automatic loop capable of performing smart synthesis of new catalysts. This would leverage the power of the iterative active learning process coupled with the usage of large language models like ChatGPT. In this envisioned framework, once an initial set of parameters and a target objective are defined, the system would autonomously navigate through multiple iterations of experimentation and learning. At each stage, the AI system would analyze the outcomes of past iterations, discern trends and correlations, and generate predictions for the next most promising sets of parameters to test. This is where the integration of large language models comes into play. By constantly ingesting new scientific literature and research findings in real-time, the AI system could continually expand its knowledge base and refine its understanding of the catalyst synthesis process. This continual learning would allow the system to make increasingly informed and sophisticated decisions about the direction of experimentation, effectively "learning" its way toward optimal synthesis parameters. With such a system in place, the optimization process becomes a continual, autonomous loop of learning and refining, capable of swiftly identifying optimal catalyst synthesis strategies. This could substantially accelerate the discovery of novel catalysts and streamline their deployment in addressing our global energy challenges.

## ACKNOWLEDGEMENT

This work is supported by the National Key R&D Program of China (No. 2022ZD0117501).